\documentclass[prd,reprint,preprintnumbers,nofootinbib]{revtex4-2}

\usepackage{amsfonts,bm,amssymb,euscript,array,babel}
\usepackage{mathtools}
\usepackage{amsmath}
\usepackage[amsmath]{empheq}
\usepackage{hyperref}

\newcommand{\be}{\begin{equation}}
\newcommand{\ee}{\end{equation}}

\def \bea{\begin{eqnarray}} 
\def\eea{\end{eqnarray}}
\def\bse{\begin{subequations}}	
	\def\ese{\end{subequations}}
\def\bal{\begin{align}} 
\def\eal{\end{align}}

\def\bi{\begin{itemize}} 
	\def\ei{\end{itemize}}

\def\a{\alpha}    \def\d{\delta} 
 
  \def\h{\eta} 
  \def\m{\mu}
\def\n{\nu} \def\o{\omega}

\begin{document}

\title{Tensor Galileons as Lovelock theories}

\author{Georgios Karagiannis}
\email{Georgios.Karagiannis@irb.hr}
\affiliation{Division of Theoretical Physics, Rudjer Bo\v skovi\'c Institute, Bijeni\v cka 54, 10000 Zagreb, Croatia}

\begin{abstract}
    We review the construction of Galileon interactions involving a single two-column mixed-symmetry tensor of arbitrary degree in flat spacetime of arbitrary dimensions, in a reverse engineering spirit. By defining an appropriate Riemann-like tensor for each of these tensor gauge fields, we show that the theories constructed in the spirit of Lovelock's theory of Gravity correspond to the well-known Galileon theories from the literature. 
\end{abstract}
	\maketitle
\preprint{RBI-ThPhys-2021-2}
\section{Introduction}
When it comes to constructing interactions in gauge theory, there are some general requirements that have to hold. First of all, the action should exhibit invariance under all (or part of) the gauge symmetry of the kinetic term. Considering the example of a 2-form gauge field $B_{\m\n}$, the corresponding gauge transformation is 
\be \label{gauge trafo}
\d B_{\m_1\m_2}=\partial_{[\m_1}a_{\m_2]}
\ee
for an arbitrary vector gauge parameter $a_{\m}$.

A second requirement is that the theory should not suffer from Ostrogradsky ghosts \cite{Ostrogradsky:1850fid}, which would imply that the Hamiltonian is not bounded from below. In general, it is easy to check whether this is the case or not by examining the corresponding field equations. More specifically, if the field equations do not contain terms having more than two time derivatives acting on the gauge field, then the Ostrogradsky instability is avoided. In relativistic theories, this implies that there should not exist terms having more than two partial derivatives acting on the respective field. 

In the special case where there only exist terms with exactly second derivatives, an additional symmetry appears. For the 2-form, this symmetry has the form
\be \label{galileon trafo}
B_{\m_1\m_2}\mapsto B_{\m_1\m_2}+v_{\a\m_1\m_2}x^\a,
\ee
where $v_{\a\m_1\m_2}$ is a constant totally antisymmetric tensor and $x^\a$ is the position vector. These special theories were originally studied for scalar fields in \cite{Nicolis:2008in}, where they were also coined as Galileons (see also \cite{Deffayet:2009mn}). Since then, Galileons have been the subject of extensive study and generalizations. In Ref. \cite{Deffayet:2010zh} the Galileon theories for $p$-form gauge fields of even degree were constructed, while a further generalization to mixed-symmetry tensor fields was carried out in \cite{Chatzistavrakidis:2016dnj}. 

In the case of $p$-forms, these interactions generally contain an even number of field appearances (the only counter-example is the 4-form cubic vertex discussed in \cite{Deffayet:2017eqq}) and it was already noticed in \cite{Deffayet:2010zh} that odd degree forms (like the Maxwell field) do not admit gauge invariant Galileon interactions that are not total derivatives. To be more precise, this observation concerns only the gauge invariance of the Lagrangian and does not refer to the equations of motion. Indeed, there has been an example of a 3-form theory with a Lagrangian that is not gauge invariant, but with gauge invariant field equations that exhibit the Galilean symmetry \cite{Deffayet:2016von}.

This latter observation was extended to the mixed-symmetry tensor case of \cite{Chatzistavrakidis:2016dnj}, where it was shown that tensors with odd \emph{total} degree also fall into this category. 
What is more, a special case of Galileon interactions for mixed-symmetry tensors of equal-length two-column Young tableau type exists and these can also contain an odd number of fields. For example, these interactions exist for the scalar field and for the graviton, corresponding to the odd vertices in the Galileon theory of \cite{Nicolis:2008in} (like the cubic scalar vertex of the DGP model \cite{Dvali:2000hr}) and in the Lovelock theory \cite{Lovelock:1971yv} (like the Gauss-Bonnet term) respectively. 

The aim of this Letter is to review the construction of Galileon theories under a different perspective. This is inspired by a formal analogy between the free field equations of an arbitrary gauge field and the linearized Einstein equation in Gravity. This analogy becomes manifest upon the definition of a ``Riemann'' tensor for the gauge field in question. In the example of the 2-form, this leads to the rewriting of its Maxwell-like kinetic term in a form akin to the Einstein-Hilbert term. Then, one can use this Riemann tensor to build higher-derivative vertices that are formally analogous to the ones contained in Lovelock theory of Gravity. The resulting theory matches precisely the well-known 2-form Galileon of \cite{Deffayet:2010zh}.

Finally, we generalize this approach to arbitrary mixed-symmetry tensor gauge fields. The corresponding higher-derivative vertices turn out to coincide with the well-known Galileon interactions constructed in  \cite{Deffayet:2010zh}, in the case of $p$-forms, and in \cite{Chatzistavrakidis:2016dnj}, for two-column mixed-symmetry tensors.

\section{The 2-form Lovelock theory}
As we already mentioned, the key ingredient of our construction is an appropriate ``Riemann'' tensor for the 2-form field. We shall define this tensor by
\be \label{Riemann}
R_{\m_1\m_2\m_3\n_1}:=\partial_{\n_1}\partial_{[\m_1}B_{\m_2\m_3]}\,,
\ee
in a way similar to the definition of the standard Riemann tensor for the linearized graviton.
This object is manifestly invariant under the gauge transformations \eqref{gauge trafo} and satisfies the differential and algebraic Bianchi identities $\partial_{[\m_1}R_{\m_2\m_3\m_4]}{}_{\n_1}=0=\partial_{[\n_1}R^{\m_1\m_2\m_3}{}_{\n_2]}$ and $R_{[\m_1\m_2\m_3\n_1]}=0$. 
Throughout this work, spacetime indices are raised/lowered using the Minkowski metric and its inverse. As usual, the trace of the Riemann tensor \eqref{Riemann} corresponds to the ``Ricci tensor'' $R_{\m_1\m_2}\equiv \h^{\m_3\n_1} R_{\m_1\m_2\m_3\n_1}$, which reads as
\be \label{trace}
R_{\m_1\m_2}=\frac{1}{3}\,\Box B_{\m_1\m_2}+\frac{2}{3}\partial^\a \partial_{[\m_1}B_{\m_2]\a}\,.
\ee
Unlike the standard symmetric Ricci tensor in Gravity, this is an antisymmetric rank-2 tensor. The obvious implication is that its traceless, i.e. the ``Ricci scalar'' $R\equiv \h^{\m_1\m_2} R_{\m_1\m_2}$ vanishes identically. That is, the Riemann tensor defined by \eqref{Riemann} turns out to be double-traceless.
Moreover, one can observe that the standard kinetic term for the 2-form can also be written as
\be 
\mathcal{L}_{(2)}\propto \d^{\m_1\dots\m_3}_{\n_1\dots\n_3}B_{\m_1\m_2}R^{\n_1\n_2\n_3}{}_{\m_3},\qquad D\geq 3\
\ee
where $\d^{\m_1\dots\m_3}_{\n_1\dots\n_3}$ is the generalized Kronecker delta.
Indeed, variation w.r.t. $B$ yields the vanishing of the Ricci tensor \eqref{trace} as an Euler-Lagrange equation. In this way, we have written the standard 2-form equations of motion in an ``Einstein'' form and its Maxwell-like kinetic term in a way that strongly resembles the linearized Einstein-Hilbert term. 

It is then natural to ask whether there exists a cubic interaction for the 2-form, in analogy to the Gauss-Bonnet term for the graviton. To this end, one could propose the following cubic vertex 
\be \label{cubic vertex}
\mathcal{L}_{(3)}\propto\d^{\m_1\dots \m_5}_{\n_1\dots \n_5}B_{\m_1\m_2}R^{\n_1\n_2\n_3}{}_{\m_3}R_{\m_4\m_5}{}^{\n_4\n_5},
\ee
which would be potentially admissible in $D\geq 5$. However, this term is identically zero \footnote{In a previous version of this manuscript, we mistakenly claimed that this cubic term corresponds to a new cubic Galileon interaction. We are grateful to C. Deffayet for pointing us out that this is not the case.} in full agreement with the results of \cite{Deffayet:2017eqq,Deffayet:2016von} that a cubic (and, in general, an odd vertex) Galileon term for a 2-form gauge field cannot exist.

Following the above recipe, the next term is quartic in $B$ and nontrivial in $D\geq 7$:
\be 
\mathcal{L}_{(4)}\propto\d^{\m_1\dots \m_7}_{\n_1\dots \n_7}B_{\m_1\m_2}R^{\n_1\n_2\n_3}{}_{\m_3}R_{\m_4\m_5\m_6}{}^{\n_4}R^{\n_5\n_6\n_7}{}_{\m_7}.
\ee
This term is manifestly invariant under the gauge transformation \eqref{gauge trafo}, being constructed from the gauge invariant Riemann tensor \eqref{Riemann}. In addition, it is easy to see that this term is not a total derivative and that its contribution to the Euler-Lagrange equations will have only second partial derivatives acting on $B$ (since the Riemann tensor satisfies the differential Bianchi identities). 
Therefore, this quartic interaction is of Galilean type and one can see that it corresponds to the quartic vertex of \cite{Deffayet:2010zh} up to a total derivative. 

Completing the theory is now straightforward; in analogy to the linearized Lovelock theory of Gravity, the higher-derivative vertices can be built by appropriately contracting the 2-form field with specific combinations of the Riemann tensor.  These combinations have the same form as the Lovelock invariants for the graviton and the full spectrum of interactions will be given by
\be \label{2form interactions}\begin{split}&\mathcal{L}_{(2k)}\propto \d^{\m_1\dots \m_{4k-1}}_{\n_1\dots \n_{4k-1}}B_{\m_1\m_2}R^{\n_1\n_2\n_3}{}_{\m_3}\\
&\qquad\times \prod_{r=1}^{k-1}R_{\m_{4r}\m_{4r+1}\m_{4r+2}}{}^{\n_{4r}}R^{\n_{4r+1}\n_{4r+2}\n_{4r+3}}{}_{\m_{4r+3}}
\end{split}\ee 
These interactions are nontrivial in $D\geq 4k-1$ and correspond to the even Galileon vertices constructed in \cite{Deffayet:2010zh}, up to a total derivative term.

\section{Generalization to mixed-symmetry tensors}
The discussion in the previous Section can be easily generalized to arbitrary two-column mixed-symmetry tensors of even total degree. As we already mentioned, it was recently shown in \cite{Chatzistavrakidis:2016dnj} that a gauge field $\o^{\m_1\dots\m_p}{}_{\n_1\dots\n_q}$, with $p+q$ being even and $p\geq q$, always admits Galileon interactions containing an even number of field appearances. Furthermore, there also exist odd vertices in the special case of $p=q$. 

Such a gauge field corresponds to an irreducible $(p,q)$ Young tableau representation of the general linear group. This implies that its local components are subject to the index symmetries 
\be \begin{split}\label{index symmetries}
\o_{\m_1\dots\m_p\n_1\dots\n_q}&=\o_{[\m_1\dots\m_p][\n_1\dots\n_q]}\,,\\
\o_{[\m_1\dots\m_p\n_1]\dots\n_q}&=0\,,
\end{split}
\ee
as well as the additional symmetry $\o_{\m_1\dots\m_p\n_1\dots\n_q}=\o_{\n_1\dots\n_q\m_1\dots\m_p}$ in the case where $p=q$.
In addition, its free theory should be invariant under the more involved gauge transformation \cite{Hull:2001iu}
\be \label{gauge trafo MS}
\begin{split}
\d\o^{\m_1\dots\m_p}{}_{\n_1\dots\n_q}=&\mathcal{P}_{(p,q)}(\partial^{[\m_1}a^{\m_2\dots\m_p]}{}_{\n_1\dots\n_q}\\
&\qquad\qquad+\partial_{[\n_1}b^{\m_1\dots\m_p}{}_{\n_2\dots\n_q]}),
\end{split}
\ee
where $a$,$b$ are arbitrary reducible mixed-symmetry tensor gauge parameters and $P_{(p,q)}$ is the projection into the index symmetries \eqref{index symmetries} of a $(p,q)$ Young tableau.

Just like before, the Riemann tensor relevant for our construction will correspond to a $(p+1,q+1)$ Young tableau and will be given by
\be 
R^{\m_1\dots\m_{p+1}}{}_{\n_1\dots\n_{q+1}}:=\partial_{[\n_1}\partial^{[\m_1}\o^{\m_2\dots\m_{p+1}]}{}_{\n_2\dots \n_{q+1}]}\,.
\ee
Note that Riemann-like tensors of this form have been extensively used in the setting of duality, see e.g. \cite{Bekaert:2002dt,deMedeiros:2002qpr}.
This tensor is invariant under the gauge transformation \eqref{gauge trafo MS} and satisfies the two differential Bianchi identities $\partial^{[\m_1}R^{\m_2\dots\m_{p+2}]}{}_{\n_1\dots\n_{q+1}}=0=\partial_{[\n_1}R^{\m_1\dots\m_{p+1}}{}_{\n_2\dots\n_{q+2}]}$ by definition. In addition, one finds an algebraic Bianchi identity and the vanishing of its $(q+2)^{\text{th}}$ trace:
\be \begin{split}
    R_{[\m_1\dots\m_{p+1}\n_1]\n_2\dots\n_{q+1}}&=0\,,\\
    \h^{\m_1\n_1}\dots\h^{\m_{q+2}\n_{q+2}} R_{\m_1\dots\m_{p+1}\n_1\dots\n_{q+1}}&=0\,.
\end{split}
\ee
The first identity is an irreducibility condition that accompanies any tensor of Young type $(p+1,q+1)$, while the trace condition is somewhat trivial, in the sense that it implies the vanishing of the trace of a fully antisymmetric rank-$(p-q)$ tensor. 

Moving on, we can see that the free equation of motion for the $(p,q)$ gauge field can be obtained by variation of the kinetic term
\be 
\begin{split}\mathcal{L}_{(2,p,q)}\propto \,\d^{\m_1\dots \m_{t+1}}_{\n_1\dots \n_{t+1}}\,&\o_{\m_1\dots\m_p}{}^{\n_1\dots\n_q}\\
&\hspace{-0.5cm}\times R^{\n_{q+1}\dots\n_{t+1}}{}_{\m_{p+1}\dots\m_{t+1}}\,,
\end{split}
\ee
where we have denoted by $t\equiv p+q$ the total degree of $\o$. 
The above expression can be seen as the analogue of the linearized Einstein-Hilbert term. The sole difference is that, in our context, it contains the irreducible $(p,q)$ gauge field $\o$ instead of the linearized graviton.

Constructing the full spectra of interactions follows precisely our reasoning in the last Section. It is in fact straightforward to generalize \eqref{2form interactions} into
\be \begin{split}\label{lovelock general}
&\mathcal{L}_{(2k,p,q)}\propto \,\d^{\m_1\dots \m_{(t+2)k-1}}_{\n_1\dots \n_{(t+2)k-1}}\,\o_{\m_1\dots\m_p}{}^{\n_1\dots\n_q}R^{\n_{q+1}\dots\n_{t+1}}{}_{\m_{p+1}\dots\m_{t+1}}\\
&\times \prod_{r=1}^{k-1}(R_{\m_{(t+2)r}\dots \m_{(t+2)r+p}}{}^{\n_{(t+2)r}\dots \n_{(t+2)r+q}}\\
&\times R^{\n_{(t+2)r+q+1}\dots\n_{(t+2)(r+1)-1}}{}_{\m_{(t+2)r+p+1}\dots\m_{(t+2)(r+1)-1}}),
\end{split}
\ee
where the interaction $\mathcal{L}_{(2k,p,q)}$ is nontrivial in spacetime dimensions $D\geq (t+2)k-1$.  Like before, these vertices always contain an even number of field appearances and the differential Bianchi identities on the Riemann tensor force the Euler-Lagrange equations to exhibit Galilean invariance. For a two-column mixed-symmetry tensor, the corresponding transformation that generalizes \eqref{galileon trafo} reads as \cite{Chatzistavrakidis:2016dnj}
\be 
\o_{\m_1\dots\m_p\n_1\dots\n_q}\mapsto \o_{\m_1\dots\m_p\n_1\dots\n_q}+v_{\a\m_1\dots\m_p\n_1\dots\n_q}x^{\a},
\ee
where $v$ is a constant totally antisymmetric tensor. Moreover, we can see that \eqref{lovelock general} reduces to the 2-form theory \eqref{2form interactions} for $p=2$ and $q=0$. 

The vertices \eqref{lovelock general} are constructed for any tensor of degree $(p,q)$ and correspond to the already known Galileon interactions of \cite{Deffayet:2010zh,Chatzistavrakidis:2016dnj}. Moreover, it is well-known that odd Galileon vertices exist, with the most prominent examples being the scalar Galileon ($p=q=0$) and the Lovelock theory of Gravity ($p=q=1$). These special odd vertices were proven in \cite{Chatzistavrakidis:2016dnj} to exist for any tensor gauge field in an equal-length Young tableau representation, i.e. with $p=q$. In the language of Riemann tensors, these theories take the simple form
\be \begin{split}\label{lovelock general p=q}
\mathcal{L}_{(n,p)}\propto \,&\d^{\m_1\dots \m_{(p+1)n-1}}_{\n_1\dots \n_{(p+1)n-1}}\,\o_{\m_1\dots\m_p}{}^{\n_1\dots\n_p}\\
&\times\prod_{r=1}^{n-1}R_{\m_{(p+1)r}\dots \m_{(p+1)r+p}}{}^{\n_{(p+1)r}\dots \n_{(p+1)r+p}}
\end{split}
\ee
and one can see that the even vertices contained in this expression coincide with the respective vertices of \eqref{lovelock general}. In this sense, \eqref{lovelock general p=q} corresponds to an enhancement of possible interactions for equal-length tensor fields.

\section{Conclusions}
Working in flat spacetime of arbitrary dimension, we have reviewed the construction of Galileon higher-derivative interactions containing a single gauge field. In general, this field can be in any two-column Young tableau representation, with the sum of the lengths of these columns being an even number. 

The key observation underlying our approach is a formal analogy between the free theories of these fields and linearized Gravity, using which the kinetic terms for these fields can be rewritten in a way that strongly resembles the linearized Einstein-Hilbert term. In each case, this is achieved by defining an appropriate Riemann-like tensor that satisfies the standard differential Bianchi identities, as well as an algebraic Bianchi identity and a (trivial) trace condition. Then, the free field equations can be written as the vanishing of the first trace of this tensor, in the same way that the linearized Einstein equation can be written as the vanishing of the Ricci tensor.

Subsequently, we constructed higher-derivative interactions in the spirit of Lovelock's theory, the fundamental constructing blocks of each vertex being the analogues of the Lovelock invariants constructed by specific combinations of these Riemann tensors.
For any choice of gauge field, these interactions are manifestly invariant under the free theory gauge transformations and exhibit the corresponding Galilean symmetry. As such, they are found to reproduce the well-known literature results on $p$-form \cite{Deffayet:2010zh} and mixed-symmetry tensor Galileons \cite{Chatzistavrakidis:2016dnj}.

\section*{Acknowledgments}
 We are grateful to C\'edric Deffayet for  correspondence. We would like to thank Athanasios Chatzistavrakidis and Peter Schupp for helpful discussions, and for collaboration in earlier work related to the subject of this paper. This work is supported by the Croatian Science Foundation Project ``New Geometries for Gravity and Spacetime'' (IP-2018-01-7615).

\end{document}